\pdfoutput=1

\documentclass{aa}

\usepackage{graphicx}
\usepackage[varg]{txfonts}
\usepackage{amsmath}
\usepackage{verbatim}
\usepackage{latexsym}
\usepackage{multirow}
\usepackage{fixltx2e}
\usepackage{color}
\usepackage{rotating}

\begin{document}

\title{Connecting the Wilson depression to the magnetic field of sunspots}

\author{B. L\"optien\inst{1}
\and A. Lagg\inst{1}
\and M. van Noort\inst{1}
\and S.~K. Solanki\inst{1,2}}

\institute{Max-Planck-Institut f\"ur Sonnensystemforschung, Justus-von-Liebig-Weg 3, 37077 G\"ottingen, Germany
\and School of Space Research, Kyung Hee University, Yongin, Gyeonggi, 446-701, Republic of Korea}

\date{Received <date> /
Accepted <date>}

\abstract {In sunspots, the geometric height of continuum optical depth unity is depressed compared to the quiet Sun. This so-called Wilson depression is caused by the Lorentz force of the strong magnetic field inside the spots. However, it is not understood in detail yet, how the Wilson depression is related to the strength and geometry of the magnetic field or to other properties of the sunspot.}
{We aim to study the dependence of the Wilson depression on the properties of the magnetic field of the sunspots and how exactly the magnetic field contributes to balancing the Wilson depression with respect to the gas pressure of the surroundings of the spots.}
{Our study is based on 24 spectropolarimetric scans of 12 individual sunspots performed with Hinode. We derived the Wilson depression for each spot using both, a recently developed method that is based on minimizing the divergence of the magnetic field, and an approach developed earlier that enforces an equilibrium between the gas pressure and the magnetic pressure inside the spot and the gas pressure in the quiet Sun, thus neglecting the influence of the curvature force. We then performed a statistical analysis by comparing the Wilson depression resulting from the two techniques with each other and by relating them to various parameters of the sunspots, such as their size or the strength of the magnetic field.} 
{We find that the Wilson depression becomes larger for spots with a stronger magnetic field, but not as much as one would expect from the increased magnetic pressure. This suggests that the curvature integral provides an important contribution to the Wilson depression, particularly for spots with a weak magnetic field. Our results indicate that the geometry of the magnetic field in the penumbra is different between spots with different strengths of the average umbral magnetic field.}
{}

\maketitle

\section{Introduction}
The Wilson depression $z_{\rm W}$ \citep{1774RSPT...64....1W} is the difference in height of unity optical depth between a sunspot and the quiet Sun. This depression is caused by the strong magnetic field of the spot. The magnetic pressure reduces the gas pressure within the spot. In addition, the magnetic field quenches the convection, leading to a less effective heat transport, and thus, to a lower temperature and to a lower opacity within the spot. Therefore, the Wilson depression should strongly depend on the strength and the geometry of the magnetic field.

The properties of the magnetic field of a sunspot depend to a large extend on its size. Larger spots exhibit a stronger magnetic field and a lower temperature in the umbra \citep[e. g.][]{1992SoPh..141..253K,2002SoPh..207...41L,2012A&A...541A..60R,2014SoPh..289.1477S,2014ApJ...787...22W,2014A&A...565A..52K,2015A&A...578A..43R}. This suggests, all else being equal, that the Wilson depression should be larger for spots with a lower brightness in the umbra or a larger area. \citet{1974SoPh...35..105P} estimated the Wilson depression geometrically when the sunspot approaches the limb (the so-called Wilson effect) and indeed, measured a shallower average Wilson depression for small spots ($z_{\rm W} = 1000$~km) than for large spots ($z_{\rm W} = 2100$km). Unfortunately, this method suffers from radiative transfer effects and it also assumes that the spot is roughly unchanged during its passage from disk center to limb. This leads to a large range of values of the Wilson depression inferred using this method between different studies \citep[from 400~km to more than 2000~km, see][]{2003A&ARv..11..153S}. Hence, the Wilson effect can only poorly constrain the Wilson depression. Alternatively, the Wilson depression can be estimated by evaluating the horizontal force balance between the sunspot and the surrounding quiet Sun \citep{1993A&A...270..494M,1993A&A...277..639S,2004A&A...422..693M}. However, this method requires assuming by how much the curvature force contributes to the force balance. This contribution might be different for different spots and so, this method cannot be used to study the dependence of $z_{\rm W}$ on spot parameters. Recently, \citet{2019A&A...626A.102A} derived a Wilson depression of up to 600~km for one sunspot by using a spectropolarimetric inversion that is based on neural networks.

Here, we derive the Wilson depression using a recently developed method that is based on minimizing the divergence of the magnetic field vector \citep{2018A&A...619A..42L}. This method is qualitatively based on an approach originally suggested by \citet{2010ApJ...720.1417P}. We infer the Wilson depression from 24 observations of 12 individual spots made with the Hinode spacecraft. We also infer the geometry of the magnetic field and the properties of the horizontal force balance within the individual spots. We then perform a statistical analysis in order to study the connection between the Wilson depression and the properties of the magnetic field of sunspots.

\section{Data}
\subsection{Hinode observations} \label{sect:data}
Our analysis is based on a sample of 24 observations of 12 different sunspots (see Table~\ref{tab:spots}), which were observed between 2006 and 2012 with the spectropolarimeter on the Solar Optical Telescope \citep[SOT/SP,][]{2007SoPh..243....3K,2008SoPh..249..167T,2008SoPh..249..233I,2013SoPh..283..579L} onboard the Hinode spacecraft. This instrument performs spectropolarimetric observations using the Fe~I line pair at $6301.5$~\AA \ and $6302.5$~\AA. The sunspots were observed in normal mode, with a spatial sampling of $0.16''$ per pixel.

We selected spots that exhibit only little influence of molecular lines in their umbrae since these strongly affect the inversion. These lines appear predominantly in large spots since these have lower temperatures in the umbra. Hence, our study is mostly based on smaller spots, as can be seen in Figures~\ref{fig:spot_exam1} to~\ref{fig:spot_exam4} and Table~\ref{tab:spots}. The first two columns in the table indicate the NOAA active region number and the date when the spot was observed. The next three columns list some general parameters of the spots: their heliocentric angle $\theta$, the area of the spot, and the mean magnetic field in the umbra at $\log \tau = -0.9$ ($B_{\rm av}$). The last two columns show the derived Wilson depressions, both for the divergence method ($z_{\rm W,div}$) and the pressure method after degrading it to the spatial resolution of the divergence method ($z_{\rm W,press}$).

\subsection{Deriving the atmospheric conditions}
We derived the atmospheric parameters for all spots by inverting the maps of the Stokes parameters with the spatially coupled version of the SPINOR code \citep{2000A&A...358.1109F,2012A&A...548A...5V,2013A&A...557A..24V} under the assumption of local thermodynamic equilibrium (LTE). We set three nodes in optical depth, placed at $\log{\tau} = -2.5,-0.9,0$, \citep[cf. ][]{2013A&A...557A..25T}. Using a spline interpolation, we then remapped the results of the inversion on an equidistant grid in $\log{\tau}$ ranging from $-6$ to $+1.5$ with a sampling of $0.1$.

Next we resolved the $180^\circ$ azimuthal ambiguity as described in \citet{2018A&A...619A..42L} by using the Non-Potential Magnetic Field Computation method \citep[NPFC,][]{2005ApJ...629L..69G}. 

We defined both, the inner and the outer boundary of the penumbra, by using a threshold for the continuum intensity. For the outer boundary of the penumbra, we use a threshold of 90\% of the continuum intensity level of the quiet Sun, after smoothing the continuum images with a 2D Gaussian with a standard deviation of 7~pixels. For the inner penumbral boundary, we use a threshold of 50\% of the continuum intensity level of the quiet Sun, without any smoothing. In case of a few spots, this fixed threshold falsely assigns the innermost parts of some penumbral filaments to be part the umbra. However, this is a small effect and does not influence the conclusions of this study.


\begin{table*}
\caption{Overview of the investigated spots and the derived Wilson depressions.}
\label{tab:spots}
\centering
\begin{tabular}{l l l l l l l}
\hline\hline
NOAA & Date & $\theta$ [deg] & Spot area [Mm$^2$] & $B_{\rm av}$ [G] & $z_{\rm W,div}$ [km] & $z_{\rm W,press}$ [km]\\
\hline
10923&    2006.11.14&     8&    2572&    2827&    673&    555\\
10933&    2007.01.04&    16&     874&    2252&    685&    327\\
10933&    2007.01.06&     9&     933&    2377&    657&    393\\
10933&    2007.01.07&    24&     871&    2473&    662&    438\\
10944&    2007.02.28&     2&     397&    2361&    612&    369\\
10944&    2007.03.01&    11&     379&    2384&    640&    378\\
10944&    2007.03.02&    16&     377&    2353&    633&    384\\
10953&    2007.04.28&    43&    1447&    2269&    601&    356\\
10953&    2007.04.30&    13&    1539&    2379&    610&    443\\
10953&    2007.05.02&    12&    1330&    2340&    589&    417\\
10953&    2007.05.03&    24&    1228&    2276&    598&    358\\
10953&    2007.05.04&    37&    1210&    2169&    584&    293\\
10960&    2007.06.10&    34&     265&    2252&    648&    348\\
10969&    2007.08.27&    12&     363&    2166&    660&    310\\
10969&    2007.08.28&    16&     353&    2116&    529&    287\\
11039&    2009.12.27&    44&     202&    2350&    626&    320\\
11039&    2009.12.28&    34&     177&    2058&    567&    273\\
11039&    2010.01.01&    29&     284&    2471&    659&    366\\
11041&    2010.01.26&    20&     170&    2068&    638&    305\\
11106&    2010.09.16&    27&     195&    2349&    636&    381\\
11117&    2010.10.27&    25&     522&    2093&    561&    291\\
11117&    2010.10.28&    35&     223&    2090&    625&    313\\
11363&    2011.12.06&    25&    1268&    2287&    524&    326\\
11536&    2012.07.31&    34&     124&    2181&    577&    346\\
\hline
\end{tabular}
\end{table*}

\section{Measuring the Wilson depression}
\subsection{Divergence method}
This method for deriving the Wilson depression is based on minimizing the divergence of the magnetic field vector. The approach is similar to the one of \citet{2010ApJ...720.1417P}, who determined the geometric height of the $\log \tau = 0$ surface of a small patch of the penumbra of a sunspot observed with Hinode by minimizing the divergence of the magnetic field and by ensuring force balance. We modified the approach of \citet{2010ApJ...720.1417P} by restricting the requirement on the field to be divergence-free. We then applied it to provide the Wilson depression of the entire spot, but with reduced spatial resolution. The details of our method are described in \citet{2018A&A...619A..42L}.

Our approach is based on minimizing the divergence of the magnetic field vector over the field-of-view, i. e., minimizing the following merit function:
\begin{align}
\chi^2 = \sum_{m,n} (\vec{\nabla} \cdot \vec{B})^2.
\end{align}
Here, the indices $m$ and $n$ indicate the coordinates of the image. The merit function is evaluated at a given height above the $\log \tau = 0$ surface and is minimized by shifting the magnetic field vector in height at each pixel, i. e., by adding an offset (the Wilson depression) to the height dependence of the magnetic field vector. Unfortunately, this approach cannot be applied to the entire spot at the original resolution of the data because the number of free parameters would be too high. The basic idea of our approach is now to write the merit function in Fourier space and to focus only on large spatial scales (both for the Wilson depression and for the magnetic field vector). This allows reducing the number of free parameters significantly. It leads to the following merit function:
\begin{align}
\chi^2 &= \sum_{m,n} (\vec{\nabla} \cdot \vec{B})^2\\
&= \frac{1}{N_{\rm x} N_{\rm y}} \sum_{k,l} \left | \mathcal{F} (\vec{\nabla} \cdot \vec{B}) \right | ^2\\
&\approx \frac{1}{N_{\rm x} N_{\rm y}} \sum_k^{k_{\rm max}} \sum_l^{l_{\rm max}} \left | \mathcal{F} (\vec{\nabla} \cdot \vec{B}) \right | ^2\\
&= \frac{1}{N_{\rm x} N_{\rm y}} \sum_k^{k_{\rm max}} \sum_l^{l_{\rm max}} \left | i k_{\rm x} \hat{B}_{\rm x} + i k_{\rm y} \hat{B}_{\rm y} + \mathcal{F} \left ( \frac{\partial B_{\rm z}} {\partial z} \right ) \right | ^2
\end{align}
Here, $\mathcal{F}$ indicates the Fourier transform in the two horizontal dimensions (quantities expressed in Fourier space are indicated by a hat) , $N_{\rm x}$ and $N_{\rm y}$ are the number of pixels in the $x$- and in the $y$-direction, and $k$ and $l$ indicate bins in horizontal wavenumber. After expressing the merit function in Fourier space, we neglected all terms above a maximum non-dimensional wavenumber $k_{\rm max}$ and $l_{\rm max}$. This truncation is justified since all terms in the merit function are positive, meaning that all terms have to be minimized separately. In addition, we express the Wilson depression in Fourier space, where we again consider only the terms up to this maximum wavenumber. These Fourier coefficients of the Wilson depression (up to $k_{\rm max}$ and $l_{\rm max}$) are the free parameters of the merit function, which we want to determine. Here we restrict the solution to $k_{\rm max} = l_{\rm max} = 3$, since the derived Wilson depressions are more likely to exhibit artifacts when considering higher wavenumbers \citep[see][]{2018A&A...619A..42L}.

We minimized the merit function using a genetic algorithm, which is very likely to return the global extremum of a given function. Afterwards, we defined $z_{\rm W} = 0$ as the height of the $\log \tau = 0$ surface averaged over the surroundings of the spot. Since the $\log \tau = 0$ surface is also slightly suppressed in the close surroundings of the spot, we do not include these in the definition of $z_{\rm W} = 0$. For the same reason, we exclude clearly magnetic structures such as pores etc.

When being applied to Hinode data, the Wilson depressions derived using this method exhibit an error of about 100~km \citep{2018A&A...619A..42L}. This error originates mainly from uncertainties in the inversion.

\subsection{Pressure method}
For comparison, we also derived the Wilson depression with another method that is based on requiring horizontal force balance between the sunspot and the surrounding quiet Sun at a given geometrical height \citep[e.g.,][]{1993A&A...270..494M,1993A&A...277..639S,2004A&A...422..693M}. This method relates the pressure inside the sunspot (the gas pressure plus the Lorentz force of the strong magnetic field)
with gas pressure in the quiet Sun outside the sunspot. Both should be in static equilibrium at a fixed geometric height. After making a few assumptions, such as radial symmetry of the magnetic field vector, no magnetic field in the azimuthal direction and a negligible effect of flows \citep{1977SoPh...55..335M}, this equilibrium can be expressed as
\begin{align}
P_{\rm gas}(r=a,z)-P_{\rm gas}(r,z) = B_{\rm z}^2(r,z)/8\pi + F_{\rm c}(r,z)/8\pi. \label{eq:press}
\end{align}
Here, $r=0$ refers to the center of the umbra and $r=a$ to a point in the quiet Sun. The second term on the right-hand side, $F_c$ is the curvature integral:
\begin{align}
F_{\rm c} = 2 \int_r^a B_{\rm z}(r',z) \frac{\partial B_{\rm r} (r',z)}{\partial z} dr'. \label{eq:F_c}
\end{align}
Equation~\ref{eq:press} can be used to infer the Wilson depression of a sunspot by comparing the total force (gas pressure plus magnetic pressure and curvature force) inside the sunspot with the pressure stratification of a reference model atmosphere of the quiet Sun (which extends downwards at least to the depth of the Wilson depression).

The derived Wilson depression depends on the curvature integral, which cannot directly be inferred from the observations (as it has to be determined at a fixed geometrical height). Commonly, one assumes $F_{\rm c} = 0$ \citep[see e.g.,][]{2004A&A...422..693M}. We applied this method by extracting the magnetic field and the pressure at $\log \tau = 0$ and by using a horizontal average of a quiet-sun region from a 3D MHD simulation of \citet{2015ApJ...814..125R} as a reference atmosphere.

As shown in \citet{2018A&A...619A..42L}, the pressure method underestimates the Wilson depression by 110\--180~km when assuming $F_{\rm c} = 0$, at least for the sunspot considered there. The main reason to apply both methods to the chosen sunspots is that by comparing the Wilson depression estimated independently by the two methods the curvature integral $F_{\rm c}$ can be derived.

\section{The derived Wilson depression}

\subsection{Maps of the Wilson depression}
\begin{figure*}
\centering
\includegraphics[width=17cm]{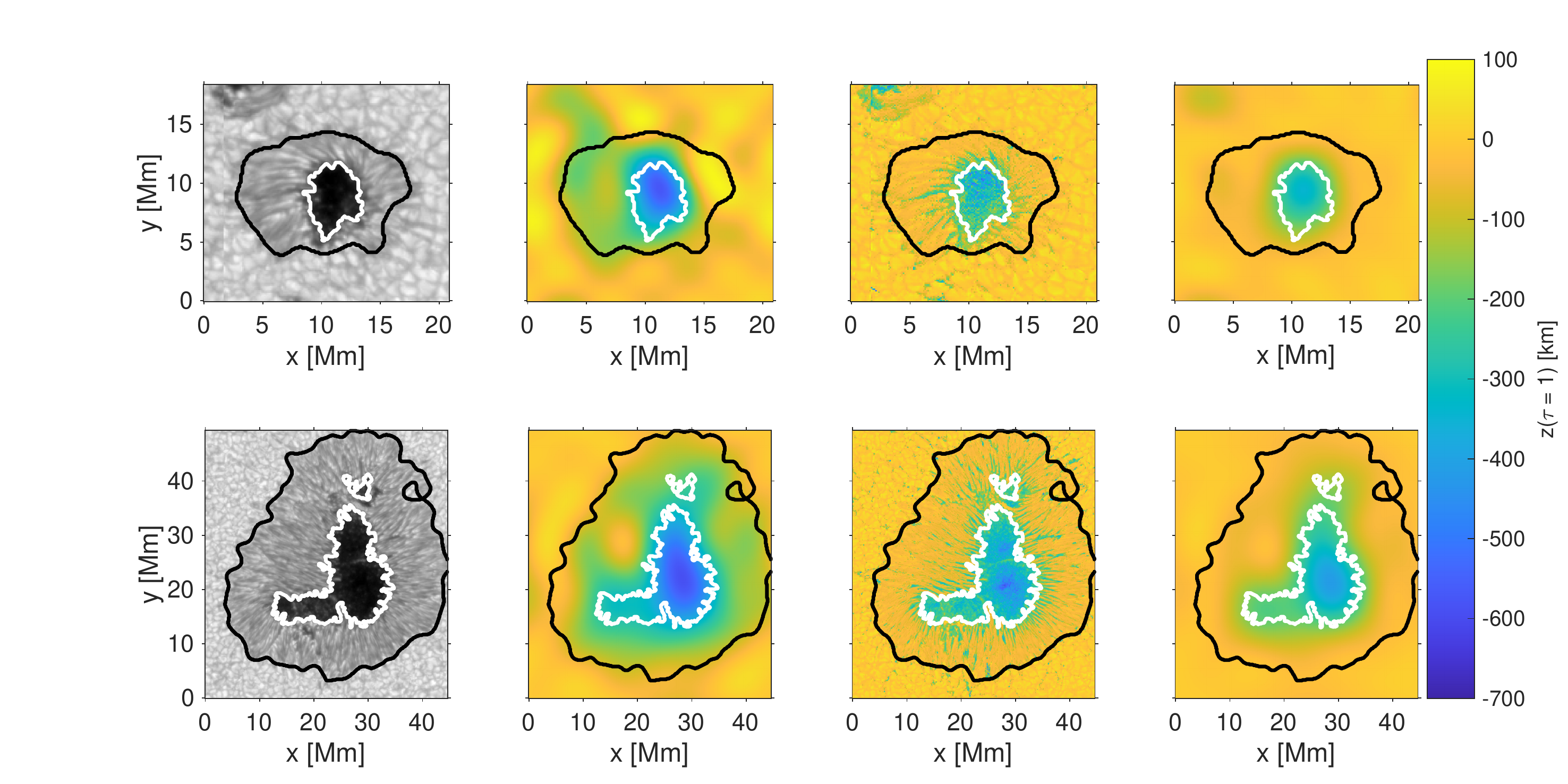}
\caption{Two of the sunspots analyzed in this study, {\it top row:} AR~11536 observed on 31 July 2012, {\it bottom row:} AR~10953 observed on 2 May 2007. We show maps of the continuum intensity and of the derived geometric height of the $\log \tau = 0$ layer across the spots (i. e., the Wilson depression, $z_{\rm W}$). {\it From left to right:} continuum intensity, $z_{\rm W}$ derived from the divergence method, $z_{\rm W}$ derived from the pressure method, and $z_{\rm W}$ derived from the pressure method degraded to the spatial resolution of the divergence method. The white and black contours indicate the inner and outer penumbral boundary, respectively, as described in the main text.}
\label{fig:spot_exam}
\end{figure*}

We derive the Wilson depression of all the spots in our sample using both, the divergence method (while setting $k_{\rm max} = l_{\rm max} = 3$) and, for comparison, also for the pressure method (both, at full resolution and after degrading to the resolution of the divergence method). Figure~\ref{fig:spot_exam} shows maps of the continuum intensity and of the resulting Wilson depressions for two sunspots, (AR~11536 observed on 31 July 2012 and AR~10953 observed on 2 May 2007). These were chosen to represent examples of small and large sunspots. The maps for all sunspots that were analyzed in this study are plotted in Figures~\ref{fig:spot_exam1} to \ref{fig:spot_exam4} in Appendix~\ref{sect:spots_all}.

There are strong differences between the Wilson depressions derived from the two methods. The divergence method gives a Wilson depression that increases smoothly from the surroundings of the spot to the center of the umbra. The region of maximum Wilson depression coincides with the shape of the umbra for both sunspots. In the penumbra, however, the Wilson depression in some cases exhibits features that are not visible in the intensity images (see e. g., AR~10969 in Figure~\ref{fig:spot_exam3}). This could be caused by neglecting the variations of the Wilson depression on small spatial scales. According to \citet{2010ApJ...720.1417P}, the geometric height of the $\log \tau = 0$ layer can vary by more than 300~km between spines and penumbral filaments. In addition, the inversion might also be inaccurate in the penumbra. The SPINOR code assumes hydrostatic equilibrium, which is not always a very accurate approximation, particularly in the penumbra. This causes errors in the stratification of the inverted atmosphere with geometric height. Localized inaccuracies of the inversion can cause large-scale errors in the Wilson depression derived using the divergence method. This is because the spatial resolution of our method is very low and because the Fourier transform is non-local (a perturbation of the signal at a given point also affects distant points after filtering in Fourier space). We note that the divergence method gives a slight Wilson depression in the close surroundings of the spot. This is probably caused by the very low spatial resolution of the divergence method.

The Wilson depression derived using the pressure method always follows closely the shape of the spot without any questionable features. This is because the pressure method treats all pixels independently. Generally, this method leads to a shallower Wilson depression than the divergence method, both, in the umbra and even more in the penumbra. It exhibits a strong decrease of the Wilson depression at the inner penumbral boundary, even after degrading the spatial resolution to that of the divergence method (see right column in Figure~\ref{fig:spot_exam} and Figures~\ref{fig:spot_exam1} to~\ref{fig:spot_exam4}). As explained in \citet{2018A&A...619A..42L}, the pressure method underestimates the Wilson depression because it only relies on the magnetic pressure resulting from the vertical magnetic field. The contributions of the (unknown) curvature integral to the Wilson depression are generally ignored by setting $F_{\rm c} = 0$. The differences are particularly strong in the penumbra because there the magnetic field is strongly inclined with respect to the vertical.

Both, the divergence and the pressure method are affected by the presence of scattered light in the Hinode data. As explained in Appendix~\ref{sect:scattered_light}, the influence of the scattered light on the divergence method is smaller than the noise level of this method (about 100~km). However, the scattered light causes the $z_{\rm W}$ provided by the pressure method to be too low.

\subsection{Statistical analysis of the Wilson depression}\label{sect:statistics}
\begin{figure*}
\centering
\includegraphics[width=17cm]{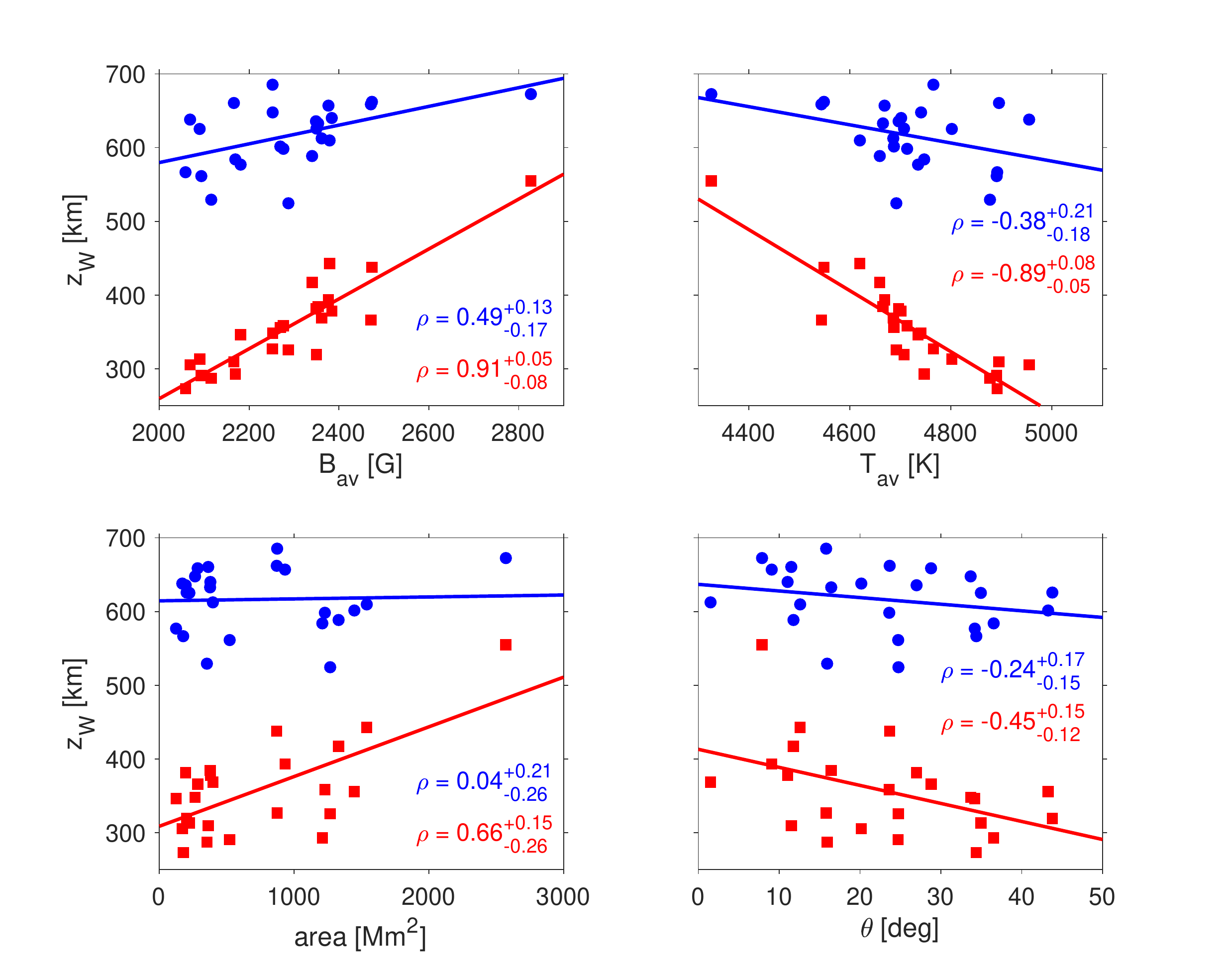}
\caption{Dependence of the maximum value of the Wilson depression $z_{\rm W}$ on various parameters of the studied sunspots. The blue circles represent the Wilson depression derived using the divergence method while the red squares depict the results from the pressure method. The solid lines indicate linear fits to the Wilson depressions. Also shown are the correlation coefficients between the Wilson depression the respective parameters of the sunspot as well as their $1\sigma$ confidence intervals. The correlation coefficients associated with the $z_{\rm W}$ derived with a given method are indicated by the corresponding color. We note that the Wilson depression inferred from the divergence method has an error of about 100~km resulting from uncertainties of the inversion. The various parameters of the sunspots are explained in more detail in the text.}
\label{fig:z_w_param}
\end{figure*}

In the next step, we evaluate systematic differences in the derived Wilson depression between the individual spots. We quantify the Wilson depression by using the maximum Wilson depression of each spot. These are listed in Table~\ref{tab:spots} and lie in the range between 500~km and 700~km (mean: 616~km, $\sigma = 44$~km) for the divergence method and between 270~km and 550~km (mean: 357~km, $\sigma = 63$~km) for the pressure method (after degrading it to the spatial resolution of the divergence method). The scatter of $z_{\rm W}$ between individual spots is consistent with the error originating from the inversion (about 100~km). The pressure method (applied with $F_{\rm c}$ assumed to be zero) always leads to a shallower Wilson depression than the divergence method (after degrading it to the spatial resolution of the divergence method, see Figure~\ref{fig:z_w_param} and Table~\ref{tab:spots}), suggesting that the curvature integral $F_{\rm c}$ is generally positive.

The inferred Wilson depressions depend significantly on some properties of the sunspots (see Figure~\ref{fig:z_w_param}). The strength of the magnetic field in the umbra has the strongest influence on $z_{\rm W}$ (see top left panel in Figure~\ref{fig:z_w_param}, here $B_{\rm av}$ is defined as the average strength of the magnetic field in the umbra evaluated at $\log \tau = -0.9$). The stronger the field, the larger the Wilson depression. This is in particular the case for the pressure method, because the magnetic pressure in the umbra directly affects the static equilibrium expressed in Eq.~\ref{eq:press}. Unfortunately, there is only one spot with a rather strong magnetic field in our sample (AR~10923, most sunspots with strong fields exhibit molecular lines in the umbra so that inversions become unreliable). Therefore, we cannot constrain the Wilson depression well for spots with strong magnetic fields. However, $z_{\rm W}$ derived from the divergence method is still correlated with $B_{\rm av}$, when not considering this sunspot (correlation coefficient $\rho = 0.43$). 

The Wilson depression also varies with the surface temperature in the umbra. Cooler spots exhibit larger values of $z_{\rm W}$ (see top right panel in Figure~\ref{fig:z_w_param}, $T_{\rm av}$ is the average temperature in the umbra at $\log \tau = 0$). This is in part because the continuum opacity is lower at lower temperatures, but predominantly because the temperature in the umbra is determined by the strength of the magnetic field in the umbra, as the heat transport by convection is more inhibited by stronger fields (the correlation coefficient between $B_{\rm av}$ and $T_{\rm av}$ is $-0.95$ in our sample). 

Interestingly, there is no significant dependence of $z_{\rm W}$ on the area of the spot for the divergence method (see bottom left panel in Figure~\ref{fig:z_w_param}), although spots with a strong magnetic field in the umbra have on average a larger area (the correlation coefficient between $B_{\rm av}$ and the area of the spot is $0.59$ in our sample). Only the pressure method clearly shows an increase of the Wilson depression with increasing spot area. We note, that there are not many large spots in our sample, though, since these are likely to exhibit molecular lines in the umbra (see Section \ref{sect:data}).

There is also a correlation between the Wilson depression and the heliocentric angle $\theta$ for both methods (see bottom right panel in Figure~\ref{fig:z_w_param}), which points to to a systematic error. This dependence is weaker for $z_{\rm W}$ obtained using the divergence method. 

To summarize, the inferred Wilson depressions depend mostly on the strength of the magnetic field in the umbra.

\section{Connection between the Wilson depression and the magnetic field}
\begin{figure*}
\centering
\includegraphics[width=17cm]{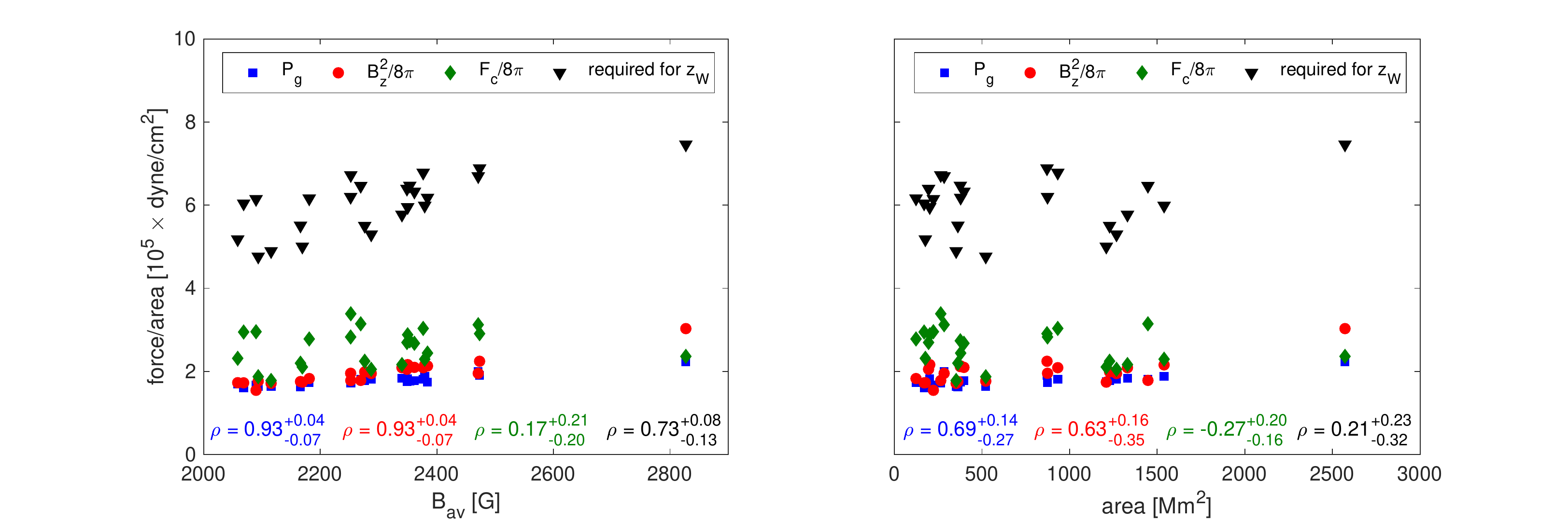}
\caption{Magnitude of the different terms contributing to the force balance given in equation~\ref{eq:press} as a function of the mean magnetic field in the umbra at $\log \tau  = -0.9 $ ({\it left}) and of the area of the spot ({\it right}). Blue squares: gas pressure ($P_{\rm g}$), red circles: magnetic pressure ($B_{\rm z}^2/8\pi$), green diamonds: curvature integral ($F_{\rm c} / 8\pi$), and black triangles: pressure required for sustaining the Wilson depression provided by the divergence method. All terms are umbral averages at $\log \tau = 0$. Also shown are the correlation coefficients between the various terms and the mean magnetic field in the umbra (left panel) or the sizes of the spots (right panel).}
\label{fig:press}
\end{figure*}

According to Eq.~\ref{eq:press}, the difference in gas pressure between the quiet Sun and the umbra is balanced by the magnetic pressure and by the magnetic curvature force. Here, we study the magnitudes of the different terms in this equation at the height of the $\log \tau = 0$ layer in the umbra. We note, however, that Eq.~\ref{eq:press} is only an approximation. Among other things, this equation does not account for advection or a twisted magnetic field. Here we ignore the contributions of these two effects.

The gas and magnetic pressure in the umbra can be directly derived from the observations when assuming hydrostatic equilibrium. The magnitude of the curvature force at $\log \tau = 0$ in the umbra can only be estimated in an indirect way, though. Since the curvature integral has to be evaluated at a fixed geometric height, it requires information below the $\log \tau = 0$ layer in the penumbra, if we stay within the photosphere in the umbra. For the same reason, we cannot directly determine the gas pressure outside the sunspot from observations. We therefore infer these two quantities in an indirect way using a reference atmospheric model. The Wilson depression inferred by the divergence method provides the geometric height we are interested in (the height of the $\log \tau = 0$ layer in the umbra). Then, we can derive the gas pressure in the quiet Sun at the same geometric height by using a reference atmosphere \citep[a horizontal average of a quiet-sun region from a 3D MHD simulation of][]{2015ApJ...814..125R}. Finally, $F_{\rm c}$ can be estimated using Eq.~\ref{eq:press}. The gas pressure in the quiet Sun that we infer indirectly from the divergence method exhibits a high scatter, significantly larger than the one of the gas pressure or the magnetic pressure in the umbra. This is because the pressure decreases almost exponentially with height. Thus, a small uncertainty in $z_{\rm W}$ leads to a large error in the gas pressure.

The differences in the inferred Wilson depressions between the two methods are not the same for all spots. As can be seen in Figure~\ref{fig:z_w_param}, the Wilson depressions derived from the two methods differ predominantly for spots with weak umbral fields, high umbral temperatures or small umbral areas. This is also visible in the different terms contributing to Eq.~\ref{eq:press}. Figure~\ref{fig:press} shows the contributions of the individual terms for spots of different sizes and for different strengths of the magnetic field. All quantities are averages over the umbra at $\log \tau = 0$. Also, we degraded the $z_{\rm W}$ provided by the pressure method to the spatial resolution of the divergence method.

The gas pressure in the quiet Sun increases strongly with depth. Hence, sunspots with a larger Wilson depression need to exhibit a higher total force (gas pressure plus magnetic pressure and curvature force) in the umbra in order to ensure force balance with the surroundings of the spot. Since $z_{\rm W}$ increases with the strength of the magnetic field in the umbra, spots with a stronger magnetic field require a higher total force in the umbra at $\log \tau = 0$ than spots with weak magnetic fields. Obviously, the magnetic pressure is larger for spots with stronger fields. The gas pressure also increases with $B_{\rm av}$ (see left panel of Figure~\ref{fig:press}). This is because the temperature in the umbra is lower for spots with strong magnetic fields. A low temperature leads to a lower opacity, which means that the formation height of the continuum is shifted towards lower layers that exhibit a higher gas pressure. For spots with a low $B_{\rm av}$, the gas and the magnetic pressure have a comparable magnitude, for spots with strong magnetic fields, the magnetic pressure exceeds the gas pressure by a factor of about $1.3$. The magnitude of the curvature integral does not change with the strength of the magnetic field. At low field strength, $F_{\rm c}$ is larger than the other contributions to the total force and at high field strengths, the magnitudes are comparable.

As discussed in the previous section, the Wilson depression does not seem to vary with spot size. Since the total force in the umbra should be in equilibrium with the gas pressure in the surroundings of the spot at a fixed geometric height, a certain value of the Wilson depression corresponds to a corresponding value of the total force. Therefore, $z_{\rm W}$ not depending on spot size implies that the total force at $\log \tau = 0$ does not change with the size of the spots, either, as observed in the right panel of Figure~\ref{fig:press}. However, there are strong variations of the different pressure terms with spot size. The gas and magnetic pressure have a comparable magnitude (the magnetic pressure is higher than the gas pressure by a factor of about $1.1$) and increase with increasing spot size. Larger spots tend to exhibit stronger magnetic fields, which cause the increase of the magnetic pressure. The higher gas pressure of large spots originates from a lower opacity, because the temperature in the umbra decreases with spot size. The stronger magnetic field and the lower temperature of large spots would therefore lead to a Wilson depression that increases with spot size. Hence, our observation of $z_{\rm W}$ not depending on spot size implies that the influence of these two effects is compensated by the curvature integral, $F_{\rm c}$, which does not appear to depend much on either spot area or umbral field strength (although it does show larger scatter than the other quantities).

For large spots, $F_{\rm c}$ contributes to the pressure balance in Eq.~\ref{eq:press} by a similar amount as the other two terms. For smaller spots, it exceeds the other terms by a factor of about $1.5$.

As already mentioned above, our derived Wilson depressions are affected by the presence of scattered light. However, the scattered light does not alter the main conclusions on the curvature integral that are described in this section (see Appendix~\ref{sect:scattered_light}).

The observed dependence of the relative influence of $F_{\rm c}$ on the area of the spot and the strength of its magnetic field suggests that there are differences in the geometry of the magnetic field between these spots. The curvature integral depends on the vertical magnetic field $B_{\rm z}(r,z)$ and on the vertical derivative $\frac{\partial B_{\rm r} (r,z)}{\partial z}$. Hence, the radial dependence of these two parameters has to vary with spot size and the strength of the magnetic field. Unfortunately, it is not possible to measure these two quantities directly. The balance between gas pressure and the Lorentz force expressed in Eq.~\ref{eq:press} is valid at a fixed geometric height. So, the magnetic field that is responsible for sustaining the Wilson depression is located in the penumbra and is located below the $\log \tau = 0$ surface. Hence, it is not accessible for spectropolarimetric observations. Only the contributions from the umbra to the curvature integral can be directly measured.

\begin{figure}
\centering
\resizebox{\hsize}{!}{\includegraphics[width=17cm]{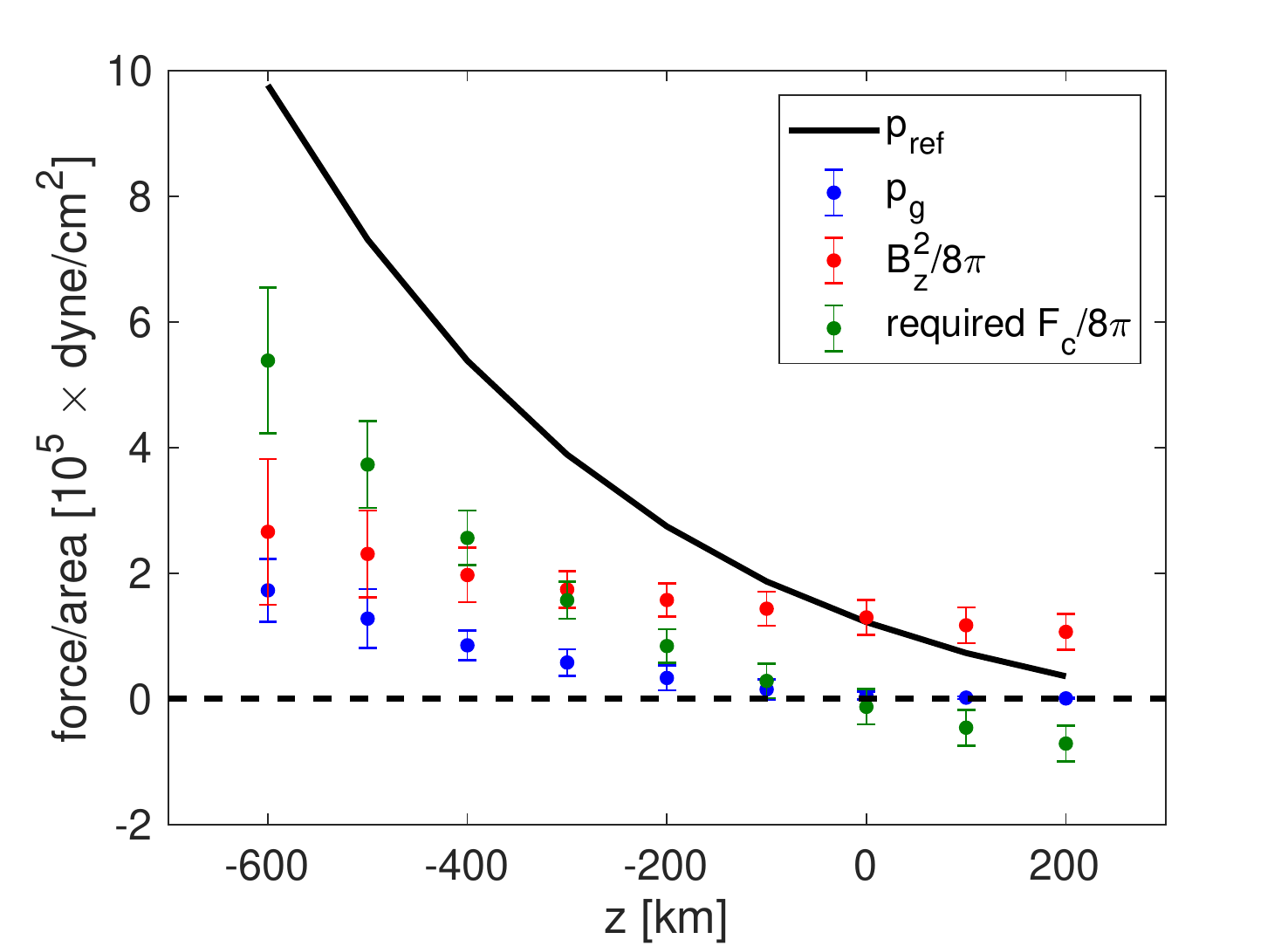}}
\caption{Height dependence of the gas pressure and the magnetic pressure given in equation~\ref{eq:press} in the umbra. Black curve: reference gas pressure ($P_{\rm ref}$) in the quiet Sun, extracted from a MHD simulation, blue: gas pressure averaged over the umbra, red: magnetic pressure averaged over the umbra ($B_{\rm z}^2/8\pi$), and green: inferred curvature integral ($F_{\rm c}/ 8\pi$). The error bars are increased by a factor of 10 in the plot for better visibility.}
\label{fig:p_height}
\end{figure}

The relative influence of the individual terms in the pressure equilibrium in Eq.~\ref{eq:press} also depends on the geometric height. Figure~\ref{fig:p_height} shows the magnitude of the gas and magnetic pressure averaged over the umbra and the strength of the curvature integral that is required for balancing the external gas pressure in the quiet Sun \citep[extracted from a quiet Sun region in a simulation by][]{2015ApJ...814..125R} averaged over all spots in our sample. All quantities were interpolated to a common grid in geometrical height using the Wilson depression inferred by the divergence method. At large depths, the curvature integral is the dominant term, but its influence decreases strongly with height, even becoming negative at around $z = 0$, meaning that it destabilizes the sunspot. At these large heights, the magnetic pressure contributes the most towards balancing the pressure deficit due to the magnetic field. It almost exactly compensates the curvature integral, suggesting that the magnetic field in the umbra becomes force-free at large heights. We note, however, that the inversion is not very accurate at $z = 0$ in the umbra as this is a few 100~km higher in the atmosphere than the $\log \tau = 0$ surface and the lines observed by Hinode/SOT/SP get relatively little contribution from this height. 

Unfortunately, it is not possible to derive the curvature integral directly, even at large heights. This is because it consists of an integral across the spot. Any systematic inaccuracies in the inversion or in the derived Wilson depression are directly summed, causing a large systematic error in $F_{\rm c}$. However, we can derive the sign of $F_{\rm c}$ for geometric heights $z > 0$. The sign is fixed by the sign of the derivative $\frac{\partial \left | B_{\rm r} (r',z) \right |}{\partial z}$. The horizontal field in the penumbra decreases with height above $\log \tau = 0$, suggesting that the curvature integral is indeed negative for $z > 0$. With the same argument, we also infer that $\frac{\partial \left | B_{\rm r} (r',z) \right |}{\partial z}\ > 0$ below $\log \tau = 0$ in the penumbra. This change of sign of the derivative $\frac{\partial \left | B_{\rm r} (r',z) \right |}{\partial z}$ at around $\log \tau = 0$ in the penumbra is also observed in numerical simulations, with the increase of $B_{\rm r}$ with height below $\log \tau = 0$ being an important ingredient for driving the Evershed flow \citep{2011ApJ...729....5R,2018ApJ...852...66S}.

\section{Discussion}
We derived the Wilson depression for a sample of 24 spectropolarimetric maps of 12 spots. Using the divergence method, we could constrain the Wilson depression to be in the range between 500-- 700~km for all spots at all the times they were observed. This is close to the lower end of values provided by the Wilson effect \citep[400 to more than 2000~km, see the discussion in][]{2003A&ARv..11..153S}. Our results are in agreement with some measurements made using the Wilson effect, such as the ones made by \citet{1972SoPh...26...52G} ($z_{\rm W} \sim 600$~km) or by \citet{1983SoPh...88...71B} ($z_{\rm W} = 500 \-- 1000$~km). Some studies based on the Wilson effect derived significantly higher values of the Wilson depression, though. \citet{1974SoPh...35..105P} measured $z_{\rm W}$ to be $1000 \-- 2100$~km with large spots having a higher Wilson depression than small spots. These large values of $z_{\rm W}$ are not consistent with the results of the divergence method. Hence, this method allows placing tight constraints on the Wilson depression.

The Wilson depression seems to be mostly affected by the strength of the magnetic field, the stronger the field, the larger the Wilson depression. This is not surprising, as the Lorentz force is stronger for larger field strengths. However, decomposing the Lorentz force into a magnetic pressure term and a curvature integral (see Eq.~\ref{eq:press}) reveals that the relative magnitude of these terms changes with the size of the spot and the strength of its magnetic field in the umbra.

This suggests that there are differences in the geometry of the magnetic field between spots of different sizes and field strengths. The curvature integral is dominated by the subsurface magnetic field in the penumbra, which is not directly accessible to observations. Hence, this issue cannot be resolved directly. Observations of the magnetic field at the surface of the penumbra, however, do suggest that there are indeed differences in the geometry of the magnetic field between small and large spots. As reported by \citet{2018A&A...611L...4J}, the magnetic field at the inner boundary of the penumbra of large spots is more inclined with respect to the vertical than in small spots. It is unclear, though, if such differences also exist below the visible surface and if they have an influence on the Wilson depression. Answering this question requires detailed numerical simulations of sunspots of different sizes and different strengths of the magnetic field. 

A knowledge of the Wilson depression is also important for measuring the total magnetic flux of the sunspot. Typically, it is measured at a fixed optical depth. This means that the magnetic field is extracted at a lower geometric height in the umbra than in the penumbra. Using our derived Wilson depressions (with the divergence method), we can estimate the influence of this effect. We compute the total magnetic flux for all spots in our sample using both, a constant optical depth ($\log \tau = 0$) and a constant geometric height ($z = 0$). The difference between the two approaches depends on the strength of the magnetic field of the spot. When taking the average over all spots in our sample, the differences are negligible. Evaluating the magnetic field at $z = 0$ results in an average magnetic flux of $(5.7\pm 0.9) \times 10^{21}$~Mx compared to $(5.4\pm 1.0) \times 10^{21}$~Mx at $\log \tau = 0$. However, the ratio between the flux at $z = 0$ and the one at $\log \tau = 0$ varies significantly between different spots, ranging from $0.79$ for AR~10923 to $1.43$ for AR~10953 observed on 28 April 2007. The reason for these large differences is related to the differences in the geometry of the magnetic field between different spots. While the strength of the vertical magnetic field decreases with height in the umbra, it increases with height in the penumbra around $\log \tau = 0$ \citep{2017A&A...599A..35J}. Exploring the details extends beyond the scope of this paper, though.

Similarly, the Wilson depression has to be taken into account when extrapolating the photospheric magnetic field to higher layers in the atmosphere. Ignoring the corrugation of the $\log \tau = 0$ layer affects the extrapolation since the magnetic field strength inside sunspots changes significantly with height. The commonly made assumption of a force-free magnetic field seems to be justified inside the umbra of sunspots, though, for geometric heights above $z = 0$.

We note, however, that there are also some constraints to the divergence method. Most of the issues addressed below were already discussed in \citet{2018A&A...619A..42L}, but are recapitulated here for the convenience of the reader. Tests performed with synthetic data performed in the above study suggest that the divergence method is reliable in the umbra, but suffers from inaccuracies and artifacts in the penumbra.

The main limitation of the divergence method are inaccuracies in the inverted atmospheres. The method requires information about the magnetic field over a broad range in height. Unfortunately, the spectral lines observed by Hinode are not ideal for this method. The Fe~I line pair is not very sensitive to the magnetic field at higher atmospheric layers and it is difficult to invert in the umbrae of sunspots with strong magnetic fields due to the presence of a myriad of molecular lines. Hence, the divergence method would benefit from the observation of multiple spectral lines, as it is planned, e. g., with the upcoming Sunrise III mission \citep[see ][for more details on Sunrise I and II]{2010ApJ...723L.127S,2017ApJS..229....2S,2011SoPh..268....1B,2011SoPh..268..103B,2011SoPh..268...35G,2011SoPh..268...57M}. Such observations can, of course, also be performed on the ground, e.g, by combining the Mg~Ib lines or the Na~ID lines with purely photospheric lines. However, the divergence method is very sensitive to stray light. Applying the divergence method to ground-based observations therefore requires a very careful treatment of scattered light in the inversion process.

Additional errors in the inverted atmospheres arise due to limitations of the inversion code itself. Inversion codes (including SPINOR) commonly assume the atmosphere to be in hydrostatic equilibrium. This affects, in particular, the stratification of the derived atmospheric parameters with geometric height. The divergence method relies on the height dependence of the magnetic field vector provided by the inversion. The assumption of hydrostatic equilibrium neglects, among other things, the influence of the Lorentz force on the stratification. In sunspots, the contribution of the Lorentz force to the vertical force balance depends strongly on position \citep{2012ApJ...744...65T}. In the umbra and the inner parts of the penumbra, the magnetic field is close to being force-free and hence, deviations from hydrostatic equlibrium are small. However, in the middle and outer parts of the penumbra, the influence of the Lorentz force increases. Most of the resulting deviations from hydrostatic equliibrium occur on the scale of the penumbral filaments and spines, which is much smaller than the spatial scales of the Wilson depression inferred with the divergence method. Nevertheless, inaccuracies in the inversion resulting from the assumption of hydrostatic equilibrium are probably at least partly responsible for the poor performance of the divergence method in the penumbra, compared to the umbra \citep[see][]{2018A&A...619A..42L}.

We note that the assumption of hydrostatic equlibrium is not necessarily contradictory to our result that the curvature integral plays an important role in balancing the Wilson depression. Our method for measuring $F_{\rm c}$ only requires the Wilson depression in the umbra, which can be determined accurately with the divergence method.

The inaccuracies in the inverted atmosphere occur predominantly on small spatial scales. Therefore, the divergence method, as we implement it, cannot be used to resolve small-scale corrugations of the $\log \tau = 0$ surface. As can be seen in Figures~\ref{fig:spot_exam1} to \ref{fig:spot_exam4}, there are artifacts present in the maps of the Wilson depression even when using $k_{\rm max} = l_{\rm max} = 3$, especially in the penumbra. Increasing the number of Fourier modes used by the divergence method does not resolve more of the fine structure of the Wilson depression, but only leads to more artifacts in the penumbra and in the surroundings of the spot. The Wilson depression in the umbra (and hence, $F_{\rm c}$) does not change significantly when increasing $k_{\rm max}$ and $l_{\rm max}$. In addition, most of the divergence of the magnetic field of a sunspot originates on spatial scales with $k=l \leq 3$ \citep[see][]{2018A&A...619A..42L}. So, the divergence method only has a limited sensitivity to $z_{\rm W}$ on smaller spatial scales. Thus, we restrict the analysis to $k_{\rm max} = l_{\rm max} = 3$.

There are also small-scale structures in the umbra, such as umbral dots or light bridges. These features exhibit a $z_{\rm W}$ that differs significantly from the surroundings, but the spatial resolution of the divergence method is by far too low for resolving the signature of these features in the derived maps of $z_{\rm W}$. However, these features could, in principle, influence the large-scale Wilson depression inferred by the divergence method. Obviously, this question cannot be addressed with the observational data used in this study, since we do not know the true Wilson depression of these spots. Tests performed using synthetic data \citep[see][]{2018A&A...619A..42L} suggest that the Wilson depression in the umbra is not significantly affected by small-scale structures (the simulations used for the tests have considerable numbers of umbral dots). Also, ignoring the corrugation of the $\log \tau = 0$ surface on small spatial scales does not have a strong influence on the large-scale magnetic field vector evaluated at a fixed geometrical height.

The above discussion suggests that the way we have implemented the divergence method is just a first step towards inferring the true geometrical height scale from spectropolarimetric inversions. A more comprehensive approach is to directly derive the absolute geometrical height scale in the inversion process, by including the Lorentz force in the force balance and ensuring $\nabla \cdot \vec{B} = 0$. This will be implemented in the FIRTEZ-dz code, that is currently being developed \citep{2019A&A...629A..24P,2019A&A...632A.111B}. Such an approach might be able to derive inverted atmospheres on a grid in geometric height that are fully self-consistent.

Another approach to obtaining more accurate Wilson depression values is via stereoscopy of sunspots in, e.g., continuum radiation. This will become possible with the Polarimetric and Helioseismic Imager \citep[PHI,][]{2019arXiv190311061S} on the Solar Orbiter mission \citep{2013SoPh..285...25M} when combined with, e. g., the Helioseismic and Magnetic Imager \citep[HMI,][]{2012SoPh..275..229S} onboard the Solar Dynamics Observatory \citep[SDO,][]{2012SoPh..275....3P}.

\begin{acknowledgements}
We are grateful to Matthias Rempel for providing the 3D MHD simulations of two sunspots. This work benefited from the Hinode sunspot database at MPS, created by Gautam Narayan. This project has received funding from the European Research Council (ERC) under the European Union’s Horizon 2020 research and innovation programme (grant agreement No 695075) and has been supported by the BK21 plus program through the National Research Foundation (NRF) funded by the Ministry of Education of Korea. Hinode is a Japanese mission developed and launched by ISAS/JAXA, collaborating with NAOJ as a domestic partner, NASA and STFC (UK) as international partners. Scientific operation of the Hinode mission is conducted by the Hinode science team organized at ISAS/JAXA. This team mainly consists of scientists from institutes in the partner countries. Support for the post-launch operation is provided by JAXA and NAOJ (Japan), STFC (U.K.), NASA, ESA, and NSC (Norway).
\end{acknowledgements}

\bibliographystyle{aa} 
\bibliography{literature} 

\appendix

\section{Maps of the continuum intensity and of the Wilson depression for all sunspots} \label{sect:spots_all}

Maps of the continuum intensity and of the Wilson depression derived with both, the divergence and the pressure method, are plotted in Figures~\ref{fig:spot_exam1} -- ~\ref{fig:spot_exam4} for all the scans studied in this paper.

\begin{figure*}
\centering
\includegraphics[width=17cm]{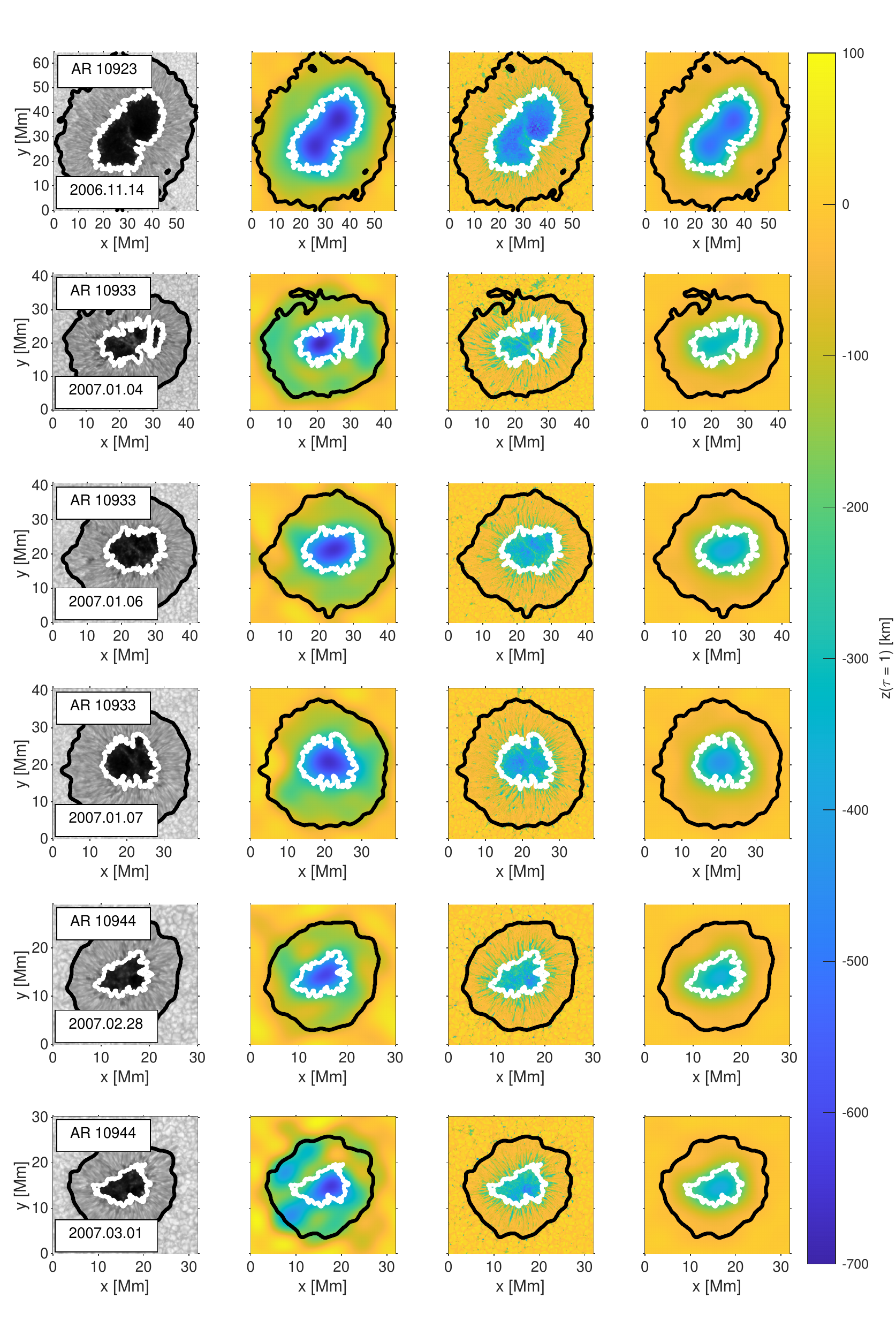}
\caption{Maps of the continuum intensity and of the derived geometric height of the $\log \tau = 0$ layer across the spots analyzed in this study (i. e., the Wilson depression, $z_{\rm W}$). {\it From left to right:} continuum intensity, $z_{\rm W}$ derived from the divergence method, $z_{\rm W}$ derived from the pressure method, and $z_{\rm W}$ derived from the pressure method degraded to the spatial resolution of the divergence method. The white and black contours indicate the inner and outer penumbral boundaries, respectively.}
\label{fig:spot_exam1}
\end{figure*}

\begin{figure*}
\centering
\includegraphics[width=17cm]{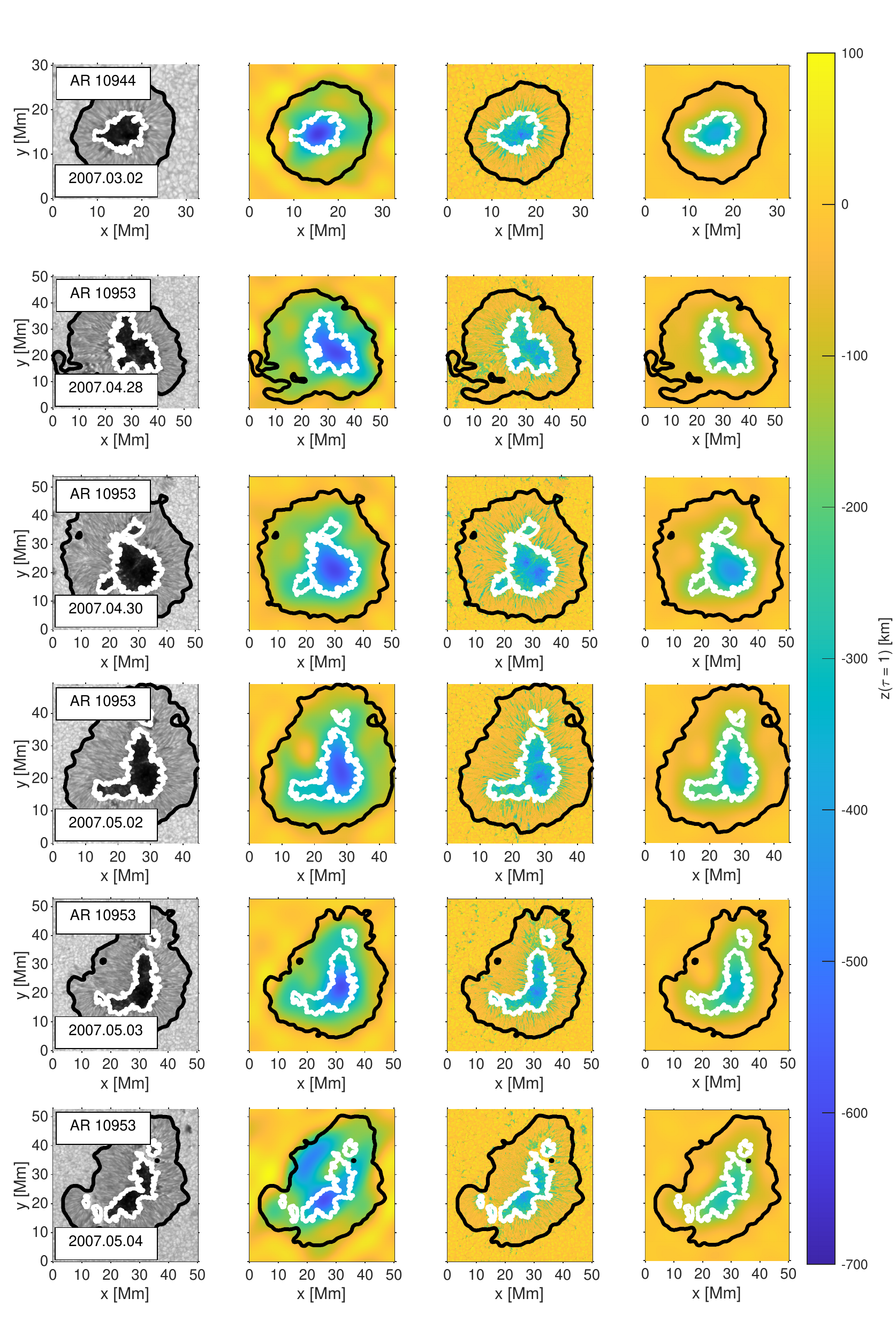}
\caption{Continuation of Figure~\ref{fig:spot_exam1}.}
\label{fig:spot_exam2}
\end{figure*}

\begin{figure*}
\centering
\includegraphics[width=17cm]{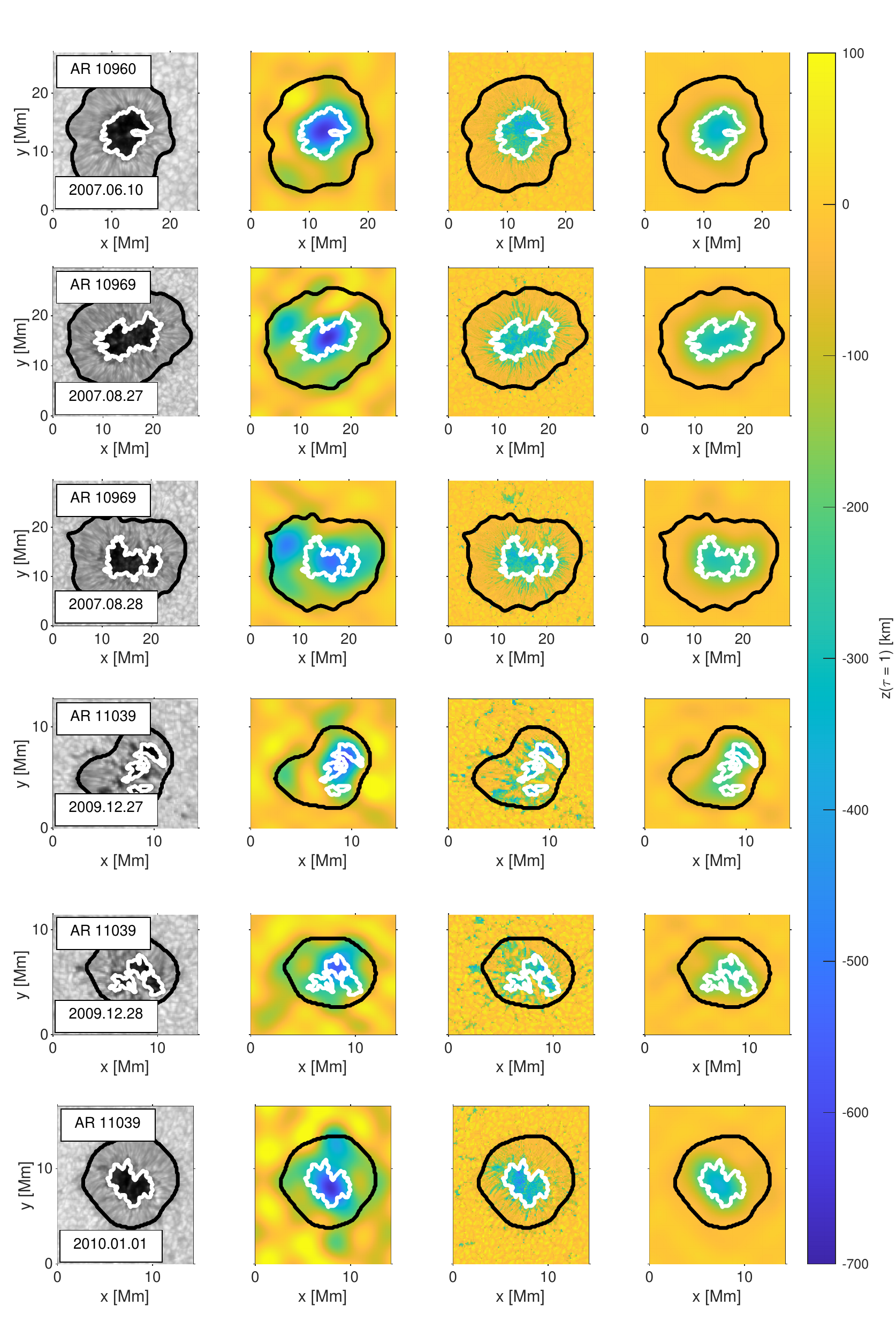}
\caption{Continuation of Figure~\ref{fig:spot_exam1}.}
\label{fig:spot_exam3}
\end{figure*}

\begin{figure*}
\centering
\includegraphics[width=17cm]{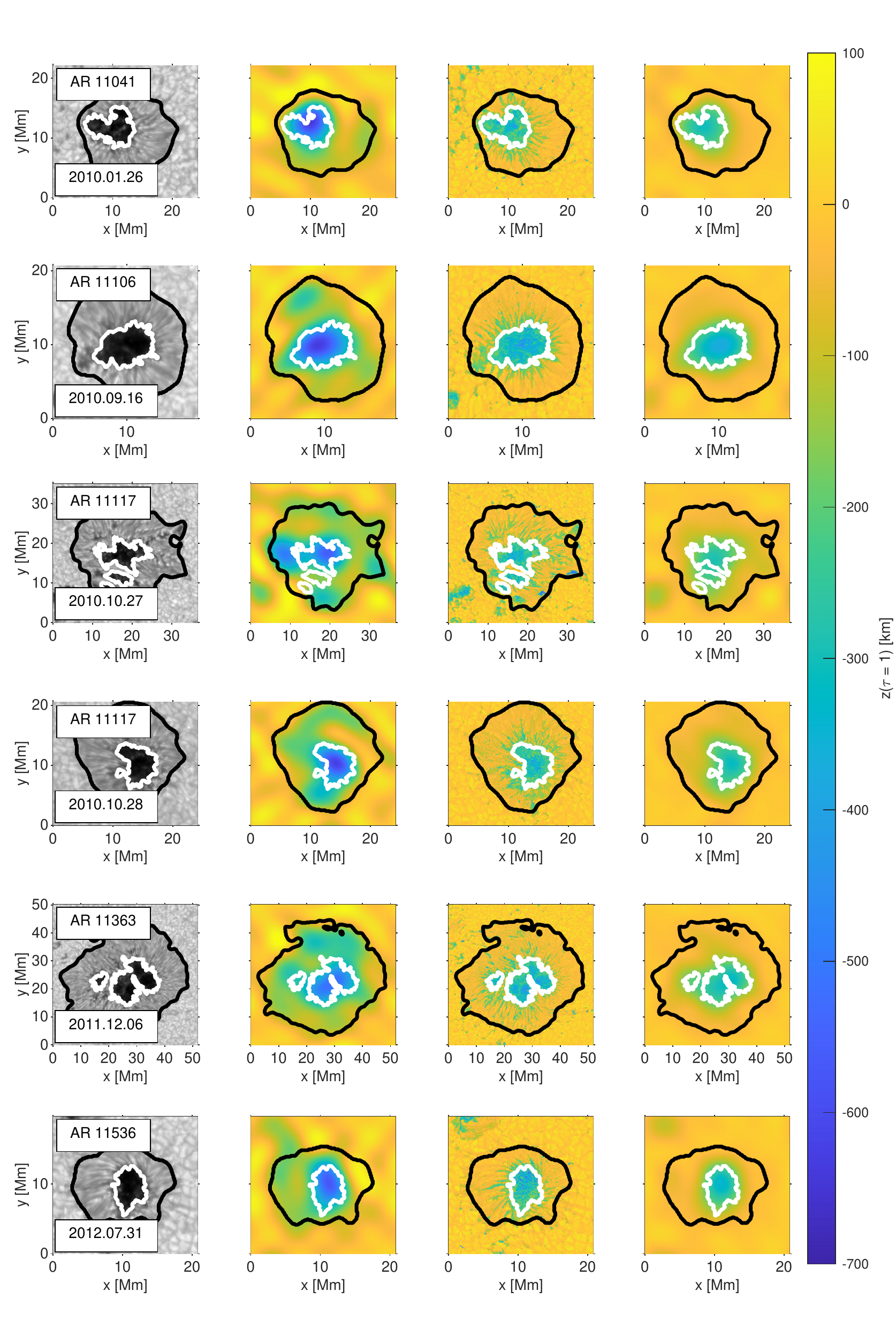}
\caption{Continuation of Figure~\ref{fig:spot_exam1}.}
\label{fig:spot_exam4}
\end{figure*}

\section{Influence of scattered light} \label{sect:scattered_light}
The SOT/SP instrument suffers from scattered light on large spatial scales \citep[about 1 \% on distances of $10''$, see ][]{2013SoPh..283..579L}. This scattered light is not taken into account by our inversions, the spatially coupled version of SPINOR only considers the influence of the point spread function. Neglecting the presence of stray light thus causes an additional error in the inverted atmospheres. Here, we test how this error affects our measurements of the Wilson depression.

We evaluate the influence of scattered light by adding additional stray light to Hinode observations of nine individual sunspots. For simplicity, we assume the scattered light to be unpolarized and to originate only from the quiet Sun. This results in the following expression for the intensity contaminated by the stray light:
\begin{align}
I(x,y,\lambda)' = (1-s) I(x,y,\lambda) + s \left < I_{\rm QS(\lambda)} \right >.
\end{align}
The factor $s$ is the magnitude of the scattered light and $\left < I_{\rm QS} \right >$ is a mean intensity profile of the surroundings of the spot. We test two different magnitudes for the additional stray light, $s=0.01$ and $s=0.02$.

The scattered light increases the intensity within the sunspots, especially in the umbra, which leads to a higher temperature in the inverted atmosphere at all optical depths in the umbra. The overestimation of the temperature implies a higher opacity, meaning that the $\log \tau = 0$ layer is shifted towards greater heights. Hence, the pressure and the density are underestimated at around $\log \tau = 0$. The higher opacity also causes lower pressure and density gradients, so that at sufficiently large heights, the pressure and the density become larger compared to the inverted atmosphere originating from data without additional stray light. The differences in the stratification of the atmosphere also affect the retrieved magnetic field vector, it decreases the strength of the magnetic field in the umbra at around $\log \tau = 0$. This is a consequence of the line weakening associated with higher temperature. Weaker lines are formed closer to $\log \tau = 0$, so that the measured Zeeman splitting of the lines now gives the same magnetic field, but at a height closer to $\log \tau = 0$.

\begin{figure}
\centering
\resizebox{\hsize}{!}{\includegraphics[width=17cm]{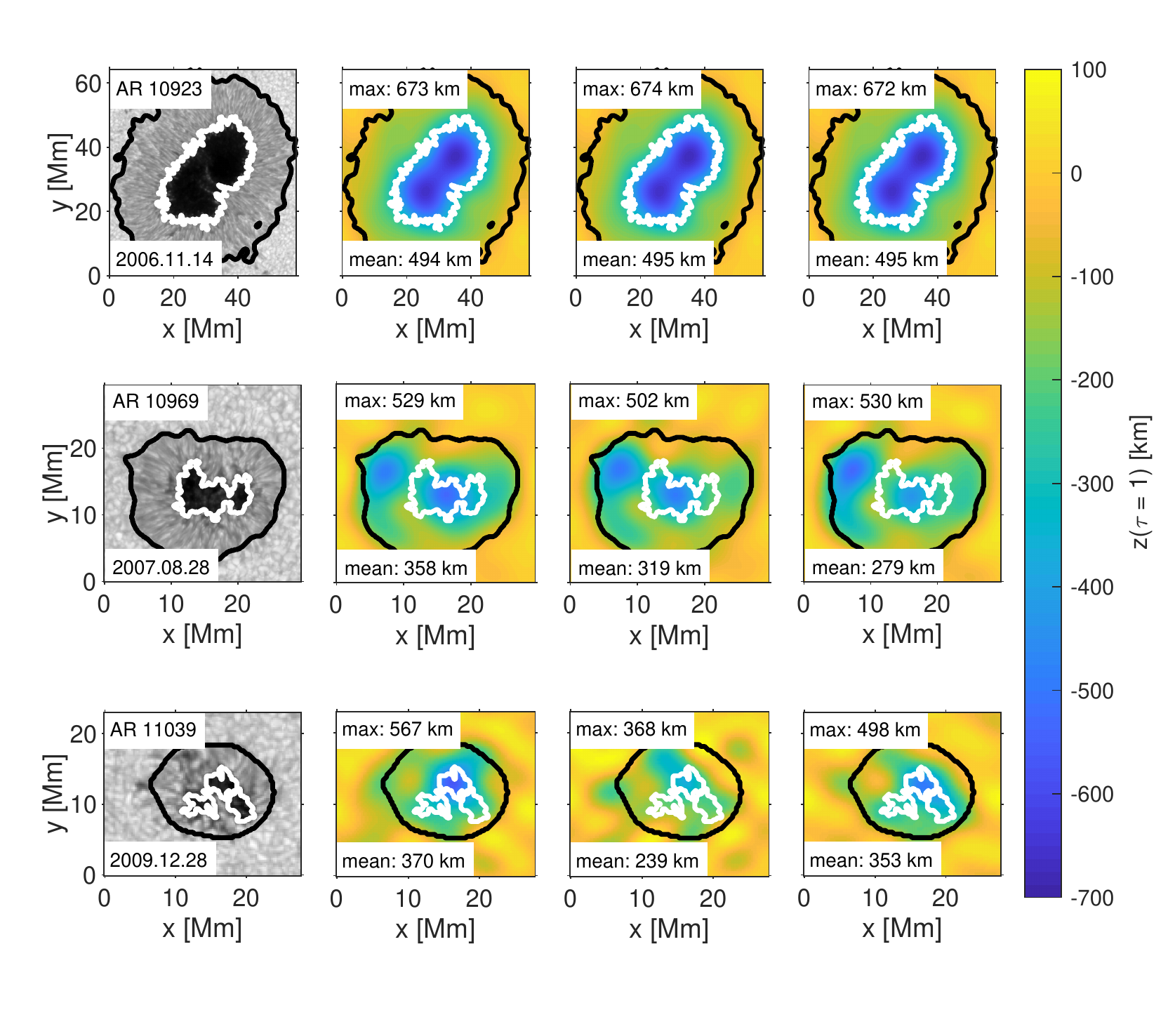}}
\caption{Comparison of maps of the geometric height of the $\log \tau = 0$ layer for selected sunspots inferred using the divergence method from inverted Stokes profiles with and without additional scattered light. {\it From left to right:} continuum intensity, $z_{\rm W}$ without additional scattered light, $z_{\rm W}$ for additional scattered light with a magnitude of 1 \% of the mean intensity of the quiet Sun, and $z_{\rm W}$ for additional scattered light with a magnitude of 2 \% of the mean intensity of the quiet Sun. The white and black contours indicate the inner and outer penumbral boundaries, respectively. The numbers give the maximum Wilson depression and its mean over the umbra.}
\label{fig:scatt_div}
\end{figure}

Adding more scattered light can affect both, the magnitude of the Wilson depression and the shape of the $\log \tau = 0$ surface. In case of the divergence method, the latter is barely altered (see Figure~\ref{fig:scatt_div}). The map of the Wilson depression of AR~10923, for example, is almost indistinguishable from the case without additional scattered light (see top row of Figure~\ref{fig:scatt_div}). Significant changes of the shape of the maps of the Wilson depression occur only for two of the spots in our sample (AR~10969 and AR~11039, see Figure~\ref{fig:scatt_div}). For a part of the penumbra of AR~10969 (west of the umbra), the divergence method measures a large Wilson depression that is comparable to the one of the umbra. This feature in the penumbra is already present in the data without additional stray light and probably is an artifact. The scattered light increases the Wilson depression of this region even more, causing it to exceed the Wilson depression in the umbra.

\begin{figure*}
\centering
\includegraphics[width=17cm]{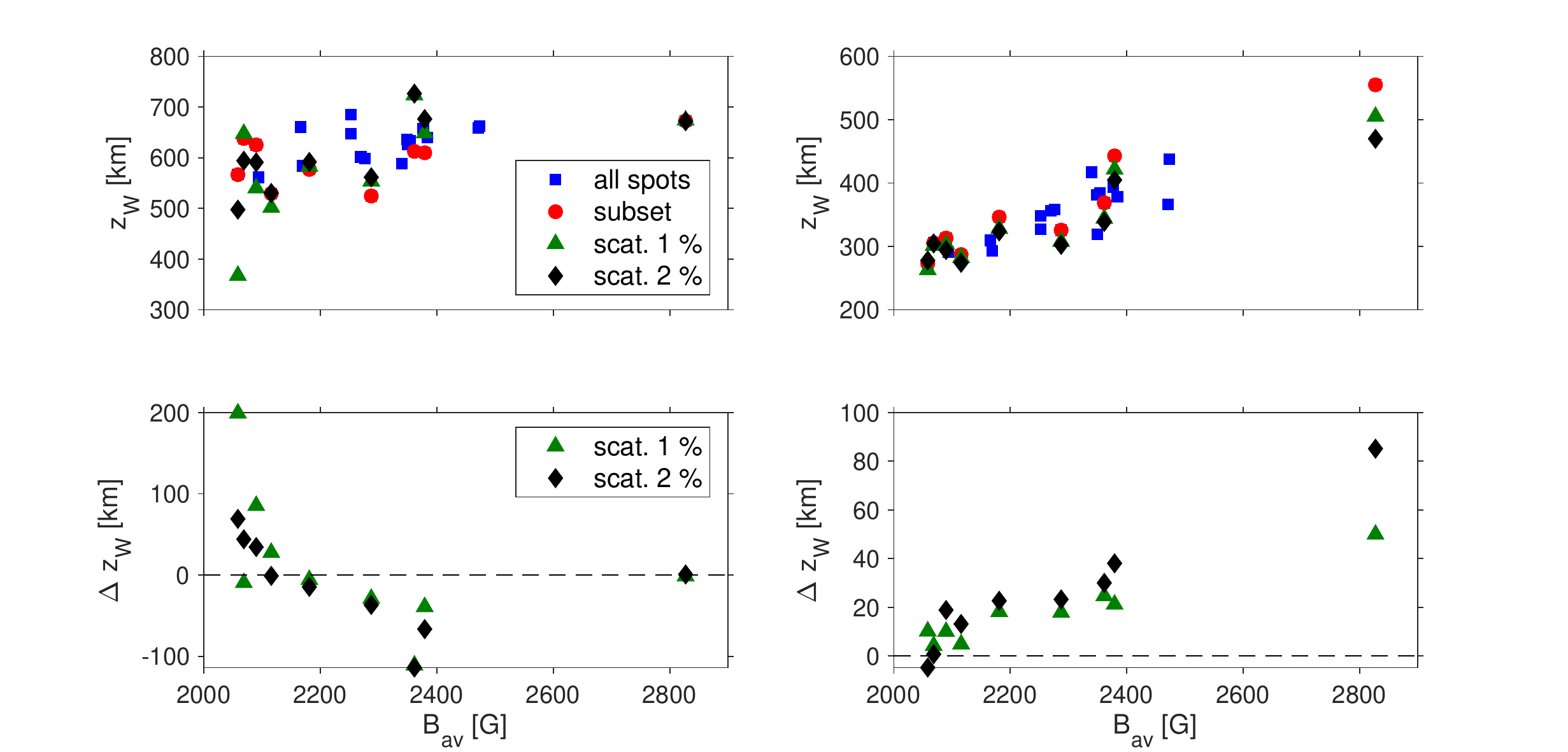}
\caption{{\it Top row:} Influence of scattered light on the maximum value of the derived Wilson depressions as a function of the strength of the mean magnetic field in the umbra at $\log \tau  = -0.9$ (derived from the inversions without any additional scattered light). We consider both, the divergence method ({\it left}) and the pressure method ({\it right}, degraded to the spatial resolution of the divergence method). The blue squares show the Wilson depression of all spots in our sample without adding additional scattered light, the red circles highlight the spots for which we evaluate the influence of the scattered light. We also show the Wilson depressions derived from data with additional scattered light. We consider two different magnitudes of the scattered light, 1 \% (green triangles) and 2 \% (black diamonds). {\it Bottom row:} Difference in the inferred Wilson depression $\Delta z_{\rm W}$ between the data without additional scattered light and with 1 \% (green triangles) or 2 \% (black diamonds) additional stray light.}
\label{fig:scatt_stat}
\end{figure*}

The additional stray light has a significant influence on the magnitude of $z_{\rm W}$ inferred by the divergence method. For some spots, the maximum Wilson depression changes by more than 100~km due to the scattered light (see top panel of Figure~\ref{fig:scatt_stat}). There are some indications that the influence of the stray light depends on the strength of the magnetic field in the umbra. For spots with strong magnetic fields, the maximum value of the Wilson depressions increases due to the scattered light (by up to $\sim 110$~km for AR~10944, $B_{\rm av} = 2361$~G). However, the sunspot with the strongest magnetic field (AR~10923) is not much affected by the additional stray light.

The additional scattered light leads to a shallower Wilson depression for almost all spots with weak magnetic fields in our sample. There is no clear trend with the magnitude of the stray light, though. For some of the spots, scattered light with $s=0.01$ affects the measured $z_{\rm W}$ more than $s=0.02$. The maximum Wilson depression of AR~11039, for example, decreases by $\sim 200$~km for $s=0.01$ but $s=0.02$ only changes it by $\sim 70$~km. The scattered light also seems to enhance the observed decrease of the derived Wilson depression with distance from disk center, but again, there is no clear trend with $s$. This strong scatter of $z_{\rm W}$ between individual spots and between the different values of $s$ could be caused by uncertainties in the inversion. The standard deviation of the difference in the maximum $z_{\rm W}$ between the results with and without additional stray light (88~km for $s=0.01$ and 57~km for $s=0.02$) is consistent with the influence of inaccuracies in the inversion \citep[about 100~km, see][]{2018A&A...619A..42L}. However, the impact of the scattered light is higher than the scatter between the maximum $z_{\rm W}$ of the individual spots ($\sigma = 44$~km) and uncertainties in the inversion should also affect the shape of the derived $\log \tau = 0$ layer, which does not seem to be the case here. Therefore, it is unclear what causes the variations in the influence of the stray light on $z_{\rm W}$ and whether the impact of scattered light is connected to $B_{\rm av}$ or $\theta$.

It is not straightforward to understand how the divergence method is influenced by scattered light. This method requires the height dependent magnetic field vector throughout the entire atmosphere. The stratification of the inverted atmosphere cannot be predicted easily from the amount of scattered light because it is also affected by the details of the inversion, such as the number and position of the nodes in optical depth that were used. These inaccuracies in the inverted atmosphere influence the divergence method in a complex manner. The Wilson depression that the divergence method measures at a given pixel depends not only on the magnetic field at this pixel, but is also affected by all other pixels within the field-of-view since in our implementation the divergence method is based in Fourier space. Therefore, the complex behavior of the inversion and of the divergence method make it hard to evaluate in detail how the Wilson depressions derived by the divergence method are affected by additional stray light. The reasons for the potential trend with $B_{\rm av}$ and for the stronger influence of $s=0.01$ remain unclear.

\begin{figure}
\centering
\resizebox{\hsize}{!}{\includegraphics[width=17cm]{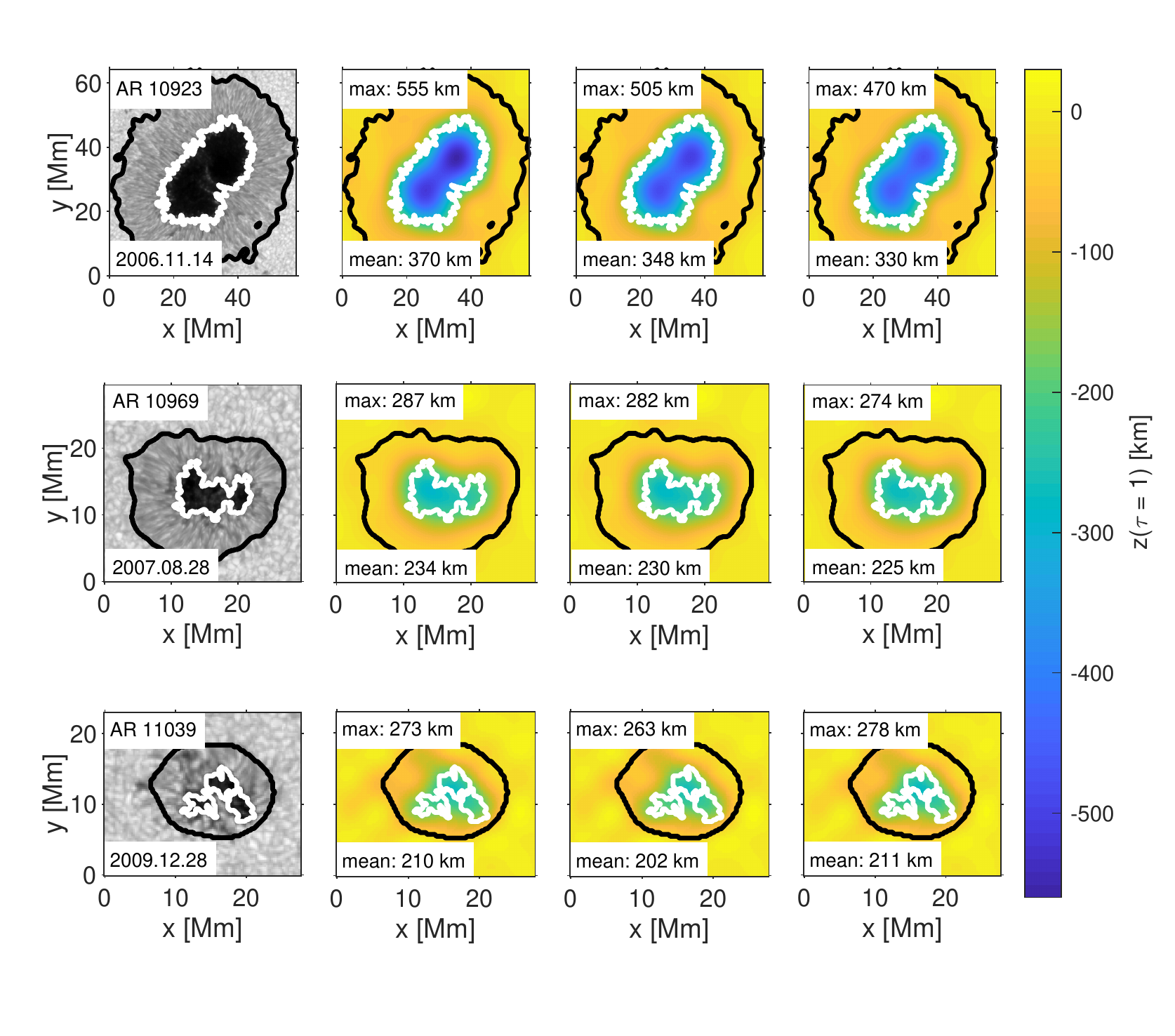}}
\caption{Same as Figure~\ref{fig:scatt_div} for the pressure method after degrading the spatial resolution to the one of the divergence method.}
\label{fig:scatt_press}
\end{figure}

The pressure method is also affected when adding additional stray light (see Figure~\ref{fig:scatt_press}). Since the scattered light causes the strength of the inverted magnetic field to decrease at $\log \tau = 0$, it also decreases the inferred Wilson depressions (with the exception of AR~11039, see below). However, the scattered light does not affect the shape of the maps of $z_{\rm W}$ inferred by the pressure method. This is because the pressure method treats all pixels in the spot independently, contrary to the divergence method, which works in Fourier space in the present implementation. AR~11039 is the only spot in our sample where the maximum Wilson depression inferred by the pressure method increases due to the scattered light (by about 5~km when using $s=0.02$). This is because there are a few pixels at the umbra penumbra boundary of this spot, where the scattered light increases the strength of the magnetic field at $\log \tau = 0$. For all other spots, the Wilson depression is decreased by scattered light. The amount by which the pressure method is influenced by the additional scattered light depends on the magnitude of the scattered light and on the strength of the magnetic field of the sunspot. The higher the magnitude of the stray light, the lower is the Wilson depression that is derived by the pressure method, while the pressure method is affected more strongly by additional scattered light for a stronger magnetic field (see bottom panel in Figure~\ref{fig:scatt_stat}). For the spot with the strongest magnetic field (AR~10923), the additional scattered light causes the maximum inferred Wilson depression (after degrading the maps of $z_{\rm W}$ to the spatial resolution of the divergence method) to decrease from 555~km to 505~km in case of $s=0.01$ and to 470~km in case of $s=0.02$. On average, the additional stray light decreases the Wilson depression measured with the pressure method by about 18~km for $s=0.01$ and by $\sim 25$~km for $s=0.02$. Hence, the pressure method is more robust regarding scattered light than the divergence method.

Adding additional stray light also affects the derived values for the curvature integral. The main reason for this is the systematic underestimation of the strength of the magnetic field at $\log \tau = 0$ due to the scattered light. The reduced magnetic pressure has to be compensated by a stronger curvature integral in order to keep the Wilson depression of the spot fixed. This is particularly the case for large spots. For example, the stray light increases the $F_{\rm c}$ of AR~10923 from $2.36\times 10^5$ dyne/cm$^2$ to $2.76\times 10^5$ dyne/cm$^2$ (for $s=0.01$) or $3.05\times 10^5$ dyne/cm$^2$ (for $s=0.02$), an increase by almost 30\%. The inferred curvature integrals are very sensitive to inaccuracies of the derived Wilson depressions, because the gas pressure in the quiet Sun decreases almost exponentially with height. Therefore, a small change in geometric height causes a large change in the gas pressure that has to be balanced, which leads to strong changes in the derived $F_{\rm c}$. The explains the larger scatter in $F_{\rm c}$ in Fig.~\ref{fig:press} compared with the gas and magnetic pressure. The scattered light leads to a lower curvature integral for the spots, though, where it causes the divergence method to underestimate the Wilson depression. Thus, the average of the curvature integral over the nine spots that we consider here is not much affected by the stray light ($2.5\times 10^5$ dyne/cm$^2$ without additional stray light, $2.3\times 10^5$ dyne/cm$^2$ for $s=0.01$, and $2.6\times 10^5$ dyne/cm$^2$ for $s=0.02$). The observed decrease of $F_{\rm c}$ with spot size cannot be caused by stray light, either. Since the main effect of the scattered light is to increase the curvature integral for large spots, the decrease of $F_{\rm c}$ with spot size would even be stronger if there was no stray light present in the Hinode data. We therefore conclude that the presence of scattered light in the Hinode data does not have a strong influence on the main conclusions of this paper.

\end{document}